\documentclass[aps,prb,preprint]{revtex4-1}
\usepackage{epstopdf}
\usepackage{graphicx}  
\usepackage{dcolumn}   
\usepackage{bm}        
\usepackage{amssymb}
\usepackage[english]{babel}
\begin{document}

\title{Pressure induced lattice expansion and phonon softening in layered $ReS_2$ }

\author{Pinku Saha$^{1}$}
\author{Bishnupada Ghosh$^{1}$}
\author{Aritra Mazumder$^{1}$}
\author{Konstantin Glazyrin$^{2}$}

\author{Goutam Dev Mukherjee$^{1}$}
\email [Corresponding Author:]{ goutamdev@iiserkol.ac.in}
\affiliation {$^{1}$National Centre for High Pressure Studies, Department of Physical Sciences, Indian Institute of Science Education and Research Kolkata, Mohanpur Campus, Mohanpur – 741246, Nadia, West Bengal, India.}

\affiliation{$^{2}$Photon Science, Deutsches Elektronen Synchrotron, 22607 Hamburg, Germany}


\begin{abstract}
We report high pressure X-ray diffraction and a detailed systematic Raman measurements on $ReS_2$ sample, which is mechanically exfoliated from a single crystal. A few new Bragg peaks are observed to emerge above 6 GPa indicating a structural transition from distorted $1T$ to distorted $1T$$^{\prime}$ in triclinic structure. The same is corroborated by appearance of new Raman modes in the same pressure range. Softening of the Raman modes corresponding to $Re$ atom vibrations are observed in the distorted $1T$$^{\prime}$ phase in the pressure range 15-25 GPa. In the same pressure range the anomalous change in the volume is found to be induced by the lattice expansion. The volume expansion is related to the sliding of layers leading to octahedral distortion and increase in octahedral volume. The sample is found to be much incompressible above 25 GPa with respect to below 15 GPa data. The same is also reflected in the Raman mode shifts with pressure.


\end{abstract}

\maketitle

\section{Introduction}
In recent times, transition metal dichalcogenides (TMDs) have shown to be promising laboratories for exploring the quantum phenomena due to their rich structural, electronic, and optical properties\cite{li08,wang12,chi14,choi14,huo14,akhinwade17}. Under pressure, layered two dimensional (2D) forms of these materials exhibit very interesting properties like superconductivity\cite{tongay14,chi18}, and charge density wave\cite{calandra11,chatterjee15}. TMDs are found to adopt different crystal structures\cite{silverma67,katzke04,enyashin11,hromadova13}. Among different structures, the most common stable phase is hexagonal (H) phase, while metastable triclinic (T) phase is also observed to be stable in few cases. In the H phase the metal atoms sit at prismatic positions, whereas in T phase they arrange themselves at octahedral positions between two chalcogen layers. Among TMD-trilayers a weak van der Waals force is present. In $ReS_2$ this force is reported to be very small,  less than 8\% with respect to that of $MoS_2$ and is caused by Peierls distortion of the low triclinic symmetry of $ReS_2$\cite{tongay14,liu15,feng15,yan17,wang18,sheremetyeva19}. This weak coupling makes bulk $ReS_2$ to exhibit monolayer behaviour\cite{tongay14}.

 Effect of high pressures in the hexagonal phase of TMD's have been extensively studied\cite{chi14,nayak14,nayak15,zhao15,duwal16,saha18,saha20}. An electronic transition from semiconducting to metallic phase is reported during isostructural transition ($2H_c$-$2H_a$) at high pressures for powder samples \cite{chi14,nayak14,nayak15,zhao15,duwal16}. In contrast, in exfoliated samples the same electronic transition is reported during strain induced structural transition from hexagonal to triclinic phase at high pressure\cite{saha18,saha20}. Very weak van der Waals interlayer interaction in distorted $1T$ $ReS_2$ demands high pressure studies, which can reduce the interlayer distances rapidly, and hence can change material properties easily\cite{hou10,tongay14,yan17,zhou17,wang18,oliva19,sheremetyeva19}. A few high pressure XRD studies on powder samples show structural transition from distorted $1T$ to $1T$$^{\prime}$ phase in the pressure range 7.7 to 11 GPa \cite{hou10,zhou17,wang18}. As these two phases differ by only layer staking order, experimental distinction by Raman study is very difficult\cite{sheremetyeva19}. From the inflection points in the ratio of Raman mode intensities, and change in the slope of Raman mode values in combination with first principle calculation, an intralayer transition followed by an interlayer transition in the range 8 - 19.6 GPa are reported by Yan {\it et al.}.\cite{yan17}. Very recent first principle study by Sheremetyeva {\it et al.} on $ReS_2$ under pressure reported two different slopes for pressure variation of Raman mode frequencies for $1T$ and $1T^\prime$ phases, respectively\cite{sheremetyeva19}. 

All the above results indeed require a systematic high pressure Raman, and XRD measurements to correlate the structural and vibrational properties. In our work, we have carried out a detailed systematic high pressure investigation on exfoliated $ReS_2$ using Raman spectroscopy and $XRD$ measurements up to about 45 GPa using diamond anvil cell (DAC).  A structural transition above 6 GPa, followed by a lattice expansion in the pressure range 15-25 GPa are observed in the new structure. In the same pressure range softening in the Raman modes corresponding to the Re atoms are observed.

\section {Experimental}
High quality layered single crystal of distorted $1T-ReS_2$ is purchased from HQ-graphene. Sample is mechanically exfoliated for the ambient and high pressure studies. We have used a piston-cylinder type $DAC$ from EasyLab Co. (UK) for our high pressure Raman spectroscopy measurements. Exfoliated samples are placed on the lower diamond anvil of culet 300$\mu$m for the collection of ambient Raman spectra. A T-301 steel gasket having a central hole of 100 $\mu$m, which is preindented to a thickness of 50 $\mu$m is used for high pressure measurements. Three different types of pressure transmitting media (PTM):  mixture of methanol-ethanol at a ratio 4:1, iso-propanol, and silicone oil are utilized for high pressure Raman measurements. For the determination of pressure few small ruby chips (approximate size 3-5 micron) are loaded along with the samples and pressure is measured from the shift of Ruby fluorescence peak\cite{mao86}. Raman spectra are collected in the back scattering geometry using a confocal micro-Raman system (Monovista from SI GmbH) using 1500 g/mm grating with a spectral resolution better than 1.2 $cm^{-1}$ using appropriate edge filter with cut off around 50 $cm^{-1}$. Sample is excited using sapphire SF optically pumped semiconductor laser of wavelength  488 nm. Raman spectra are collected with a 50X microscope objective lens (infinitely corrected) having laser spot size about 2$\mu$m. Laser power is kept constant to a maximum 15 $mW$  to avoid the heating of the sample.

X-ray diffraction measurements under ambient conditions as well as at high pressures are carried out in Petra-III P02.2 beamline using monochromatic X-ray having wavelength 0.2907 \AA. Very narrow beam of X-ray around 1.2$\times$2.3 $\mu$m$^2$ is chosen for the diffraction measurements. For high pressure $XRD$ measurements, the exfoliated samples are loaded in the symmetric anvil $DAC$ with culet flat 150 $\mu$m. In this case, neon is used as pressure transmitting medium. Small amount of gold powder is mixed with the sample, which acts as pressure marker \cite{dewaele08}. Sample to detector distance, tilt angle are calibrated using $CeO_2$ standard. Collected two dimensional diffraction images are integrated to intensity versus 2$\theta$ profile using dioptas software\cite{prescher15} and then analyzed using the CRYSFIRE\cite{shirley02} and Rietveld fitting program of GSAS\cite{toby01}.

\section{Results and Discussion}
\subsection{Characterization of the ambient sample}
For characterization of the crystal structure, we have recorded XRD pattern of the sample under ambient conditions. The XRD pattern is indexed to a distorted layered structure of triclinic symmetry with space group $P\bar{1}$. Lattice parameters of this phase are found to be $a$ = 6.581(7), $b$ = 6.389(8), and $c$ = 6.443(7) $\AA$; $\alpha$ = 118.9(3)$^\circ$, $\beta$ = 94.1(2)$^\circ$, and $\gamma$ = 106.4(4)$^\circ$ with volume ($V_0$) = 220.34(5) $\AA^3$ and  has an excellent agreement with the $1T$-phase in literature\cite{wang18,hou10,murray94,lamfers96}. The XRD pattern along with its Rietveld refinement is shown in Fig.1(a). 
Rietveld refinement is carried out using the atom positions given in literature as the starting model\cite{murray94,lamfers96}. Excellent fitting of the ambient XRD pattern of $ReS_2$ sample is evident from the figure. In a unit cell each rhenium atom is surrounded by six sulfur atoms occupying an octahedral coordination as shown in Fig.1(b). $ReS_6$ octahedra are packed together such that mono-layer of $ReS_2$ is stacked along $a$-axis(Fig.1(c,d)). Fig.1(c) shows the stacking of the layers along $a$ axis. In a unit cell there are 4 Re atoms and they are arranged in a zigzag way as seen from Fig.1(d) due to the Peierls distortion\cite{tongay14,zhou17}, which restricts the ordered stacking.

As per space group symmetry, each unit cell consists of 12 atoms corresponding to 36 phonon modes at center of the Brillouin zone. Out of 36 phonon modes, 18 are Raman active\cite{feng15,zhang17}. In Fig.2 we have shown the Raman spectrum in the spectral range 100-450 cm$^{-1}$ of the sample placed on lower diamond anvil culet flat at ambient conditions. We have observed all 18 vibrational modes, and they are listed at the right side of the Fig.2. All the mode values match very well with the literature \cite{zhang17,wang18,chenet15,yan17,tongay14}. Modes below 250 cm$^{-1}$ can be attributed to the vibrations of heavy $Re$ atoms and those above 250 cm$^{-1}$ to the vibrations of relatively light element, $S$-atoms in the unit cell. $A_{1g}$, $E_g$, and $C_p$  modes are related to the out-off-plane, in-plane, and coupled vibrations of $Re$ and $S$ atoms, respectively. The positions of the $A_{1g}-1$, $E_{g}-1$, $E_{g}-3$, and the difference between $E_{g}-1$ and $A_{1g}-1$ modes are dependent on the number of layers as per the analysis of unpolarized Raman spectra reported by Chenet {\it et al.}\cite{chenet15}. They observe that $A_{1g}-1$ mode frequency increases non-linearly from the initial value of 133.1 $\pm$ 0.1 cm$^{-1}$ corresponding to single layer upon increase of the layer number, whereas $E_{g}-1$, and $E_{g}-3$ mode frequencies decrease linearly\cite{chenet15}. We find $A_{1g}-1$ mode at 132.7 cm$^{-1}$, which is very close to the value for single layer as reported by the above study. Difference of $E_{g}-1$ and $A_{1g}-1$ modes is reported to decrease with the number of layer non-linearly\cite{chenet15}. In our study this difference is found to be 20 cm$^{-1}$, which is larger than single layer value 16.8 $\pm0.2$ cm$^{-1}$ reported by Chenet {\it et al.}\cite{chenet15}. Other studies on the layer dependence of Raman spectra\cite{quia16,yan18} show that the difference increases with the number of layers from 16.7 cm$^{-1}$, and has a value 19.9 cm$^{-1}$ for four layer configuration. Ratio of the intensities of $E_{1g}-1$ and $E_{1g}-2$ modes is found to be around 1.7 using 488 nm laser excitation, which is close to  1-2 layer as per McCreary {\it et al.}\cite{mccreary17}. It is also reported that bulk $ReS_2$ behaves like mono layer due to vibrational decoupling\cite{tongay14}. Due to the above discrepancies in the layer number identification using Raman spectra analysis, and mismatch in the mode position, and their difference values in the present study, we shall identify our exfoliated sample as a few layered sample.

\subsection{High pressure studies}
Pressure evolution of the X-ray diffraction patterns at selected pressure points at room temperature are shown in Fig.3(a). Above 6.1 GPa, a few new diffraction lines are found to appear around 2$\theta$ =  8$^\circ$. For clear visualization we have plotted the diffraction patterns at a few selected pressure points from 1.9 to 16.1 GPa in the Fig.3(b). From this figure it is evident that new diffraction lines appear as pressure increases, which are marked by black arrows. In contrast to the observation of a new single XRD peak as reported by Hou {\it et al.} and Wang {\it et al.}, we have observed 3-4 new diffraction lines as pressure increases\cite{hou10,wang18}. All the diffraction patterns above 6.1 GPa are indexed to a different set of lattice parameters in the same triclinic structure with space group $P\bar{1}$ indicating an iso-structural transition. In Fig.4(a) we have shown the Rietveld refinement fit of the new structure at 13.2 GPa. The same structure is reported as distorted $1T$$^{\prime}$ in the literature\cite{hou10,zhou17,wang18}. Lattice parameters of this structure at 6.1 GPa are found to be  $a$ = 6.243(5) $\AA$, $b$ = 6.641(7) $\AA$, $c$ = 6.465(7)$\AA$, $\alpha$ = 102.8(2)$^\circ$, $\beta$ = 103.4(4)$^\circ$, and $\gamma$ = 124.0(3)$^\circ$ with volume ($V$) = 194.46(4) $\AA^3$ and are in excellent agreement with those reported by Wang {\it et al.}\cite{wang18}. Packing of the $ReS_6$ octahedra in a unit cell are shown  in Fig.4(b). All the atoms are observed to occupy slightly different positions in the unit cell with respect to those reported by Wang {\it et al.}\cite{wang18}. The atom positions of two different structures are compared with the 1T$^{\prime}$ structure at 20.1 GPa reported by above group\cite{wang18} in the Table-I. In this structure sulfur atoms of adjacent layer are observed to penetrate inside the unit cell of mono-layer $ReS_2$ to form a triple layer contribution in a unit cell (Fig.4(c)). Mono-layers are formed in the $bc$-plane and are stacked in the direction of $a$-axis, which is evident from this figure. Rhenium atoms are found to present a zigzag chain in similar fashion as observed in the previous structure (Fig.4(d)). 

In Fig.5(a) we have plotted pressure dependence of volume till 42 GPa. Volume of this structure is found to decrease upon increasing pressure up to about 16 GPa and shows anomalous changes in the pressure range 16-25 GPa. In this pressure range a volume expansion is observed followed by the normal compression behaviour above 25 GPa. Pressure induced strain has an very important role in changing the structural as well as the electronic properties of materials\cite{polian11,jana16,saha18,saha20,saha21}. To see the response of strain in our sample, we have estimated Eulerian strain ($f_E$) and corresponding normalized pressure ($H$) using  the following equations\cite{angel01,polian11}:
\begin{equation}
H=\frac{P}{3f_E(1+2f_E)^{5/2}}, and
\end{equation}

\begin{equation}
f_E =\frac{1}{2}[(\frac{V_0}{V})^{2/3}-1];
\end{equation}
\noindent where $V_0$ is the ambient pressure value 220.34 $\AA^{3}$.
$H$ is linear with respect to $f_E$\cite{murnaghan37,birch47}:

\begin{equation}
H=K_0 + \frac{3}{2}K_0(K^{'}-4)f_E
\end{equation}
\noindent
where, $V$ is the volume at pressure $P$, $K_0$ is the bulk modulus, and $K^{'}$ is the first derivative of bulk modulus. Eulerian strain versus normalized pressure is shown in Fig.5(b). A jump  in $H$ is observed above 6.1 GPa, where the iso-structural transition occurs. In the pressure range 16-25 GPa, $H$ is almost doubled, even though $f_E$ remains almost constant indicating a deformation of the unit cell to accommodate application of pressure. Deformation of unit cell probably leads to expansion in volume in this pressure range. Above 25 GPa the reduced pressure value is found to decrease with increasing pressure. We have performed three different linear fits separately: (i) in the first phase (0-4.2 GPa); (ii) in the pressure range 6.1-16 GPa, (iii) in the pressure range 25-44 GPa, with the two later ranges are in the second phase. Fitting to the first phase reveals $K_0$ = 44(2) GPa and $K^{\prime}$ = 1.6(3), indicating almost linear and large compressibility behaviour. Reported values of  $K_0$, and $K^{\prime}$ for bulk powder samples in literature are: 23$\pm$4 GPa, and 29$\pm$8, respectively by Hou {\it et al.}\cite{hou10}; 35.6$\pm$5.2 GPa, and 10.8$\pm$2.4,respectively by Wang {\it et al.}\cite{wang18}. Bulk modulus value in the present case is found to be high with respect to Hou {\it et al.}\cite{hou10}, while it observed to be close to Wang {\it et al.} within their error limit\cite{wang18}. First group perfomed the experiment on the powder sample using alcohol mixture as PTM using 30$\times$30 $\mu$m$^2$ X-ray beam of wavelength 0.4959 $\AA$ and obtained $K_0$, and $K^{\prime}$ values by fitting P-V data to 3$^{rd}$-order BM EOS considering four data points (ambient to 6.4 GPa). They measured pressure from the ruby scale\cite{mao86}. A small error in determination of pressure would result in a large deviation in volume, and their observed $K_0$, and $K^{\prime}$ values. Indeed they reported $K_0$ to be 49$\pm$3 GPa while taking second-order BM EOS. On the other hand, other group carried out the experiment on powder sample using neon as PTM using 5 $\mu$m diameter X-ray beam of wavelength 0.4066 $\AA$, and determined $V_0$, $K_0$, and $K^{\prime}$ values by fitting P-V data to 3$^{rd}$-order BM EOS considering eight data points (1-7.7 GPa). They also estimated $K_0$, and $K^{\prime}$ from the fitting of $H vs f_E$ plot, which is calculated taking $V_0$ from the analysis of BM EOS as they do not report ambient volume of the sample. They also measured pressure from ruby. In contrast to the above studies, pressure in our study is estimated from the analysis of the XRD pattern of Au. XRD patterns of both sample and Au are collected from 2$\times$3 $\mu$m$^2$ area simultanuously that minimizes the error in the measured pressure values. We have carried out experiment on exfoliated sample, while all other experimental conditions remain similar to that of Wang {\it et al}\cite{wang18}. We have determined $K_0$ and $K^{\prime}$ from the fitting of the $H vs f_E$ curve taking $V_0$ from our study, and it produces relatively small error in the $K_0$ and $K^{\prime}$ values that ensures a very good determination of these parameters. Fitting of P-V data to 3$^{rd}$-order BM EOS yeild $K_0$ = 43.9$\pm$7 GPa, and $K^{\prime}$ = 1.4$\pm$0.2, and are slightly higher than that of Wang {\it et al}\cite{wang18}. Both estimations following two different paths in this study are found to agree very well. So one can speculate that different experimental conditions, determination of pressure using different methods, relatively small pressure range and lack of perfect determination of sample volume may result in a different values of bulk modulus and its derivative. 

 Fitting to the 1$^{st}$-range of the second phase provides $K_0$ = 40(2) GPa and $K^{\prime}$ = 5.9(6), which shows a larger pressure dependence of bulk modulus in comparison to the first phase. Interestingly above 25 GPa the fitting gives the values of $K_0$ = 109(5) GPa and $K^{\prime}$ = 2.7(6), much larger bulk modulus in comparison to lower pressures. Similar bulk modulus values are also observed in other transition metal dichalcogenides in their triclinic phase\cite{saha18,saha20}. The fitting of the Eulerian strain versus normalized pressure in the study by Wang {\it et al.} taking 8.9 GPa unit-cell volume as reference volume produced a bulk modulus of 90.1 $\pm$ 2.2 GPa and a pressure derivative of 5.1 $\pm$ 0.3, and the value of bulk modulus in their study is found to be very close to present study\cite{wang18}. We have shown the EOS fitting in Fig.5(a) taking the values of $K_0$ and $K^{\prime}$ from the fitting of $f_E$ {\it vs} $H$ plot and ambient sample volume in this present study as $V_0$. In Fig.5(c) we have shown the evolution of lattice parameters in both the phases. In the low pressure phase maximum compression is observed along $a$-axis, the stacking direction of the mono-layers, as expected due to the weak van der Waal's interaction along that direction. Compression of the $c$ axis is observed to be high with respect to $a$ axis up to 15 GPa in the second phase. Above 15 GPa $c$-axis remains almost unchanged up to 25 GPa, while $a$-axis increases slowly similar to volume. Above 25 GPa both the axes show high compressibility with respect to $b$-axis.

To understand the volume expansion behaviour in the range 16 - 25 GPa, we have plotted certain atom separations with pressure (Fig.6). The distance between different sets of atoms (Re5-Re6, Re6-S2, Re5-S3 and Re6-S3) are measured in a single unit cell (Fig.6(a \& b)), while between S2 and S3 is measured considering two sulfur atoms from adjacent layers. From top left corner of Fig.6(a) it is evident that S2 is of different mono-layer and S2-S3 distance is between side by side S2 and S3 atoms from adjacent layers. All the distances show similar behaviour, decrease up to 14.7 GPa then suddenly change their trend and increase up to 25 GPa, and followed by a decrease till 42 GPa. It is expected that an increment in the distance among Re6 and Re5, S2, and S3  atoms in the unit cell in pressure range 16-25 GPa would result in a decrease in distance between S2 and S3 atoms of the adjacent layers. Rather we find that it also increases in the same pressure range. It is only possible in the case of presence of layer sliding. Therefore possibly to accommodate the pressure compression behaviour the layers slide and it gives an impression of increased volume. Each unit cell contains four octahedra, two of them originated from Re5 atoms (Type I), and other two from Re6 atoms (Type II) surrounded by sulfur atoms. In Table-II, the octahedral volume (OV), average bond length (ABL), and distortion index (DI) in few pressure points are presented. Their values in the ambient sample are found to be very similar irrespective of the types (I \& II) of the octahedra. In 1T$^{\prime}$ structure at 6.1 GPa, the OV of the different types are observed to differ by 0.2 $\AA^3$, and ABL by only 0.006 $\AA$. But a large difference 0.018 in the DI of different types of octahedra with respect to the value 0.0008 in the ambient structure is observed. DI is defined as $\frac{1}{n}\sum_{i=1}^{n}\frac{d_i-d_{av}}{d_{av}} $, where $d_i$ is the distance from the central rhenium atom to the $i^{th}$ coordinating atom, and $d_{av}$ is the ABL\cite{baur74}. High value in the DI in Type II octahedral can be noted in Table-II, which means it is highly distorted. OV of Type II increases slowly from 19.3234 $\AA^3$ to a value 19.7951 $\AA^3$ at 14.7 GPa, while it decreases rapidly (from 19.0949 to 15.892 $\AA^3$ in the pressure range 6.1-14.7 GPa) in the case of Type I octahedra. DIs are observed to highly sensitive to pressure and reached to a maximum value at 14.7 GPa indicating highly distorted octahedra. Both the OVs are observed to be increased by a maximum 6.9\% in the pressure range 16-23 GPa, while in the same pressure range the unit cell volume is increased by 3\% as evident from Fig.5. Interestingly 20-25\% decrease in the DIs for both types of octadedra in the above pressure range are observed. From Fig.6(b) one can see S3 atom is shared by both type of octahedra in the 1T$^{\prime}$ phase, and it also evident from the Table-II that ABLs are increased by 2\% at 23 GPa with respect to 14.7 GPa. Therefore one can expect, increase in Re5-S3 or Re6-S3 distance should decrease S2-S3 distance. In contrary we observed an increase in the S2-S3 distance at above pressure region. So for the stabilization of the 1T$^{\prime}$ phase a minimization in DIs are observed due to increase in the unit cell volume, which can be accommodated by the layer sliding.  Above 23 GPa, a systematic behaviour like decrease in the octadedra volume, average bond length and pressure induced increase in the DI values are observed.

As a complementary study to $XRD$ measurements we have carried out Raman spectroscopy investigation up to about 44 GPa on exfoliated $ReS_2$. Earlier high pressure Raman investigation on single crystal $ReS_2$ using methanol-ethanol mixture as pressure transmitting medium, and 532 nm laser excitation source did report: (i)emergence of few new modes in the pressure range 8-19.6 GPa; (ii) changes in the Raman mode behaviour: change in the slope of intensity ratio of $E_g-3$, and $E_g-4$ modes\cite{yan17}. These were attributed to a phase transition starting at 8 GPa and completing at 19.6 GPa\cite{yan17}. But other Raman measurements on the powder samples do not report any of the above anomalous behaviour, even though evidence of structural transition is observed in their studies\cite{tongay14,wang18}. In Fig.7(a) we have shown the Raman spectra of $ReS_2$ from ambient to a pressure 7.7 GPa in the frequency range 250 to 550 $cm^{-1}$. These modes originate from the vibration of sulfur atoms. All the  $C_p$ Raman modes corresponding to the coupled vibration of rhenium and sulfur atoms are found to decrease in intensity with pressure and finally become undetectable above 6.3 GPa. In the same pressure range, a few new X-ray peaks are observed and the pattern is indexed to a different set of lattice parameters showing an iso-structural transition to the distorted $1T$$^{\prime}$phase. Therefore disappearance of $C_p$ Raman modes may be related to the increased distortion in the sample, which increases the disorder in Re-S coupling parameters. Vibrational modes $E_{g}-6$, and $A_{1g}-4$ of the sulfur atoms are found to survive in the whole pressure range in our study. Raman spectra in the frequency range 120-265 $cm^{-1}$ corresponding to the vibration of the rhenium atoms at the selected pressure points are shown in Fig.7(b). Above 6.3 GPa few Raman modes are found to emerge with pressure. We have indicated the new modes by black arrows at 7.7 GPa pressure spectrum in Fig.7(b). The emergence of these Raman modes itself is an evidence of phase transition. The intensity of these new modes are found to increase with pressure. New modes  are found to merge with the nearest existing modes above 24 GPa and broaden extensively above 24 GPa. From the Fig.7(b) it is evident that Raman modes are red shifted at 17.3 GPa with respect to 13.9 GPa data, and these are marked by the red arrows.

To have more insight, we have plotted pressure evolution of  Raman modes corresponding to rhenium atom vibrations in Fig.8(a), and those related to sulfur atoms vibrations in Fig.8(b). Interestingly, three different linear regions (1$^{st}$ range: 0-14, 2$^{nd}$ range: 15-25, and 3$^{rd}$ range: 25-45 GPa ) are noted in Fig.8(a). Raman mode values with pressure show change in the slope at around 14 GPa and 25 GPa. In the first range of the pressure 0-14 GPa, the modes are found to be highly sensitive to pressure. It is in agreement with the X-ray studies, which show high compressibility of the sample in the pressure range 0-16 GPa. Above 15 GPa all the modes soften with pressure up to 25 GPa, which gives rise to the negative slopes. One possible explanation of the softening of the mode values with pressure can be attributed to the decrease in bond strength, which is caused by the volume expansion induced by the layer sliding as observed in our X-ray diffraction studies. Above 18 GPa, the out-of-plane mode $A_{1g}-2$ is broaden extensively and disappear above 25 GPa. All remaining modes are found to harden with pressure above 25 GPa at a slow rate with respect to the 1$^{st}$ range of pressure. The smaller sensitivity of the Raman mode shift with pressure reveal the low compressibility of the sample above 25 GPa, which supports X-ray diffraction studies of this work. First-principles study of $ReS_2$ under pressure show pressure hardening of the Raman modes with smaller slope in $1T$$^{\prime}$ phase with respect to those of $1T$ phase\cite{sheremetyeva19}.  Though no softening of any Raman mode is observed corresponding to the sulfur-atom vibrational modes, $E_{g}-6$, and $A_{1g}-4$ modes show three different linear regions distinguished  by three different slope values. The out-of-plane vibrational mode, $A_{1g}-4$ of the sulfur atom  is found to be highly sensitive to pressure in the 1$^{st}$ range of pressure in the Raman measurements with respect to all the other modes. Modes related to rhenium-sulfur atom coupled vibrations, $C_{p}$'s are also found to be highly sensitive to pressure but disappeared above 6.3 GPa. $E_{g}-5$ mode broadens and disappears above 23 GPa. Slopes of the pressure variation of Raman modes are listed in the Table-III. 

Softening in the Raman modes in this study are observed just above the freezing point ($\sim$ 11 GPa) of methanol-ethanol mixture\cite{klotz}. For the exact confirmation of the Raman mode behaviour we have carried out Raman spectroscopy measurements using three different types of pressure transmitting media. In Fig.9, we have compared the mode evolution with pressure for a few selected pressure points in the pressure range 14.8-26.6 GPa. Inset of Fig.9(a) represents a loaded DAC at 17 GPa. From the Fig.9, it is evident that the modes softening in the pressure range 15-24 GPa observed in experiments using all different pressure transmitting media: Fig.9(a) corresponds to ethanol-methanol mixture; Fig.9(b) corresponds to iso-propanol; and Fig.9(c) corresponds to  silicone oil. A larger broadening of Raman modes are observed in Fig.9(c) compared to other figures and can be related to non-hydrostatic stress induced by silicone oil as it freezes at much lower pressure. Similar softening behaviour even with the use of different pressure media confirms that the mode softening is related to sample response to pressure only. The phonon softening can be related to anomalous volume expansion observed in the sample using neon as pressure transmitting medium. Theoretical thermodynamic calculations by Sheremetyeva {\it et al.} show that $1T$$^{\prime}$ phase is more favorable in the ambient condition with respect to $1T$ phase\cite{sheremetyeva19}. But the synthesis procedure of these single crystal materials support the stability of the $1T$ phase due to unavoidable finite strain. It can be noted that the Eulerian strain value increases rapidly up to 16 GPa to a value 0.075, then it slowly decreases to 0.06 at 25 GPa, as evident from Fig.5(b). It can be seen in Fig.8(a) that the maximum softening in the mode values is observed in $E_g-4$, which relates to the in plane vibration of the rhenium atom. Therefore one can speculate  that  growing strain starts decreasing due to the expansion in the Re5-Re6 distance, lattice parameter, and volume, which affect the in-plane lattice vibrations mostly. So, pressure induced strain has an important impact in the stabilization of the structure of $ReS_2$ similar to other exfoliated TMDs\cite{saha18,saha20}. More theoretical and experimental investigation are required to understand the exact effect of pressure induced strain on these systems.   

\section{Conclusions}
In the present study, we have carried out detailed high pressure Raman and $XRD$ investigations on exfoliated $ReS_2$ sample. Ambient sample is found to have distorted $1T$ structure. A structural transition to distorted $1T$$^{\prime}$ phase is detected from the emergence of a few Bragg peaks above 6 GPa. Lattice expansion due to the decrease in the Eulerian strain is observed in the pressure range 16-25 GPa, where all the Raman modes corresponding to rhenium atom vibrations show mode softening irrespective of the pressure transmitting medium. These observations show instability in the new structure in the above pressure range, and show systematic behaviour with pressure above 25 GPa. The volume expansion in the intermediate pressure range can be related to layer sliding to minimize the lattice strain.

\begin{acknowledgments}
GDM wishes to thank Ministry of Earth Sciences, Government of India for financial support under the project grant no. MoES/16/25/10-RDEAS. PS, and BG wish to thank DST, INSPIRE program by Department of Science and Technology, Government of India for financial support. GDM, and PS thankfully acknowledge DST-DESY project under Department of Science and Technology, Government of India for financial support to carry out the proposed experiment at DESY, Germany.  
\end{acknowledgments}

\noindent
{\bf {Author Contributions}} All authors have equal contribution. All authors reviewed the manuscript.

\section{Additional Information}

{\bf {Competing Financial Interests:}} The authors declare no competing financial interests.


\newpage
\begin{table}[h!]
	\centering
	\caption{\label{tab:table1} Fractional occupancies of atoms in the unit cell for two structures compared with Wang {\it et al.}\cite{wang18}.}
	
	\begin{tabular}{cccccc}
		\hline
		Pressure (GPa)& Structure& Atom &      x &   y     &     z \\
		\hline
		&& S1 &  0.208(7)   &   0.242(5)     &  0.388(8)    \\
		&& S2 &      0.277(7) &   0.785(8)    &     0.387(5) \\
		Ambient&Distorted& S3 &     0.760(5)  &   0.278(7)     &    0.125(6) \\
		&1T& S4 &     0.700(6) &   0.739(5)     &     0.119(8) \\
		&& Re5 &      0.490(7) &   0.059(5)     &     0.250(6)\\
		&& Re6 &      0.504(8) &   0.512(7)     &     0.298(5) \\
		\hline
		&& S1 &      0.092(4) &   0.889(7)    &     0.317(6) \\
		&& S2 &      0.085(7) &  0.754(8)     &     0.830(8) \\
		13.2&Distorted& S3 &      0.312(6) &   0.704(7)     &    0.568(5) \\
		&1T$^{\prime}$& S4 &      0.486(7) &   0.779(8)     &    0.206(6) \\
		&& Re5 &      0.287(6) &   0.295(7)     &     0.890(5) \\
		&& Re6 &     0.314(4) &   0.288(5)     &    0.063(5) \\
		\hline
			&& S1 &      0.01602 &   0.78922    &     0.31310 \\
		&& S2 &      0.06250 &  0.80417     &     0.85492 \\
		20.1&Distorted& S3 &      0.38012 &  0.71832      &    0.63886 \\
		&1T$^{\prime}$\cite{wang18}& S4 &      0.44306 &   0.74693     &    0.18126 \\
		&& Re5 &      0.28652 &   0.28046     &     0.48797 \\
		&& Re6 &     0.32067 &   0.30300     &    0.06016 \\
		\hline
	\end{tabular}
\end{table}

\newpage
\begin{table}[h!]
	\centering
	\caption{\label{tab:table2} The derived parameters from the analysis of the XRD patters using GSAS\cite{toby01} at a few pressure points.}
	
	\begin{tabular}{ccccc}
		\hline
		Pressure (GPa) \& structure & Type of octahedra  & OV ($\AA^3$)& ABL ($\AA$) &DI \\
		 		\hline
		Ambient, 1T	&Type I& 17.8390& 2.4293& 0.03770\\
		&Type II & 17.9318 & 2.4235& 0.03688\\
		\hline
		6.1, 1T$^{\prime}$& Type I& 19.0949& 2.5305 &0.02874\\
		&Type II& 19.3234& 2.5241& 0.04683\\
		\hline
		14.7, 1T$^{\prime}$	& Type I& 15.8920& 2.4095& 0.07568\\
		&Type II & 19.7951& 2.5428& 0.05651\\
		\hline
		23, 1T$^{\prime}$	& Type I & 16.5819& 2.4511& 0.05077\\
		&Type II&21.1620& 2.6020& 0.04217\\
		\hline
		31.3, 1T$^{\prime}$ & Type I& 15.7427& 2.3966 & 0.08317\\
		&Type II& 19.1420 & 2.5110 &0.05340\\
		\hline
		
	\end{tabular}
\end{table}

\newpage
\begin{table}[h!]
	\centering
	\caption{\label{tab:table3} Slopes of different Raman modes in three linear region.}
	
	\begin{tabular}{cccc}
		\hline
		Modes&Ambient-14 GPa&16-24 GPa&25-45 GPa\\
		\hline
		$A_{1g}-1$& 0.12&-0.43 & 0.11\\
		$A_{1g}-2$& 0.96&-0.2 & \\
		$E_{g}-1$&0.79&-0.06&0.21\\
		$E_{g}-2$&0.93&-0.22&0.71\\
		$E_{g}-3$&0.99&-0.14&0.80\\
		$E_{g}-4$&1.19&-0.86&0.34\\
		$E_{g}-5$&1.16&0.6&\\
		$E_{g}-6$&1.25&0.6&0.69\\
		$C_{p}-4$&2.92&&\\
		$C_{p}-5$&3.20&&\\
		$C_{p}-7$&2.70&&\\
		$C_{p}-8$&3.60&&\\
		$A_{1g}-4$& 3.22&2.50 & 2.06\\
		\hline
	\end{tabular}
\end{table}

\begin{figure}[h!]
	\centering
	\includegraphics[width=0.7\textwidth]{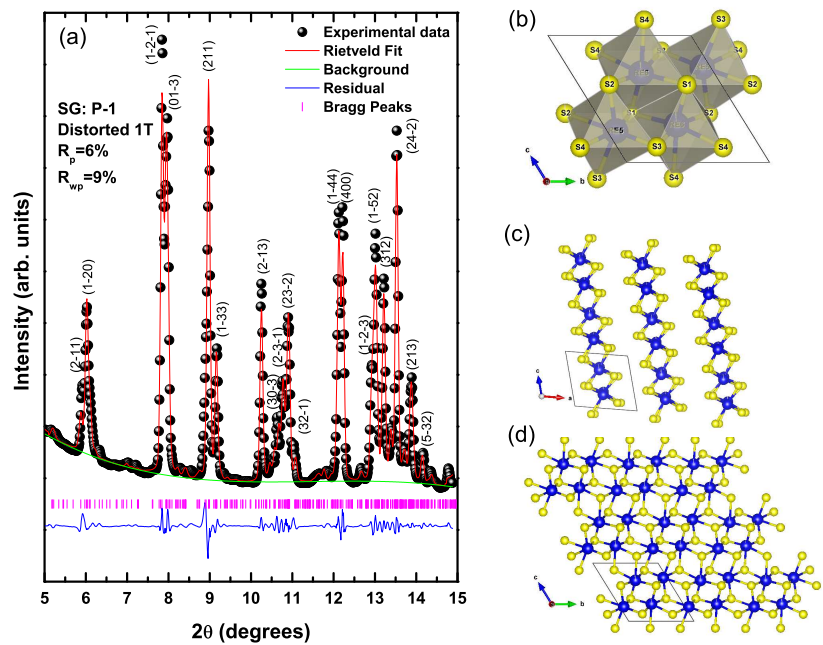}
	\caption{ (a) Rietveld refinement of the ambient x-ray diffraction pattern of $ReS_2$ sample. (b) The view of the octahedra in a unit cell of distorted 1T $ReS_2$. (c) The cross sectional view of layered 1 T $ReS_2$. (d) Top view of the mono-layered distorted 1T $ReS_2$.}
\end{figure}
\newpage
\begin{figure}[h!]
	\centering
	\includegraphics[width=0.9\textwidth]{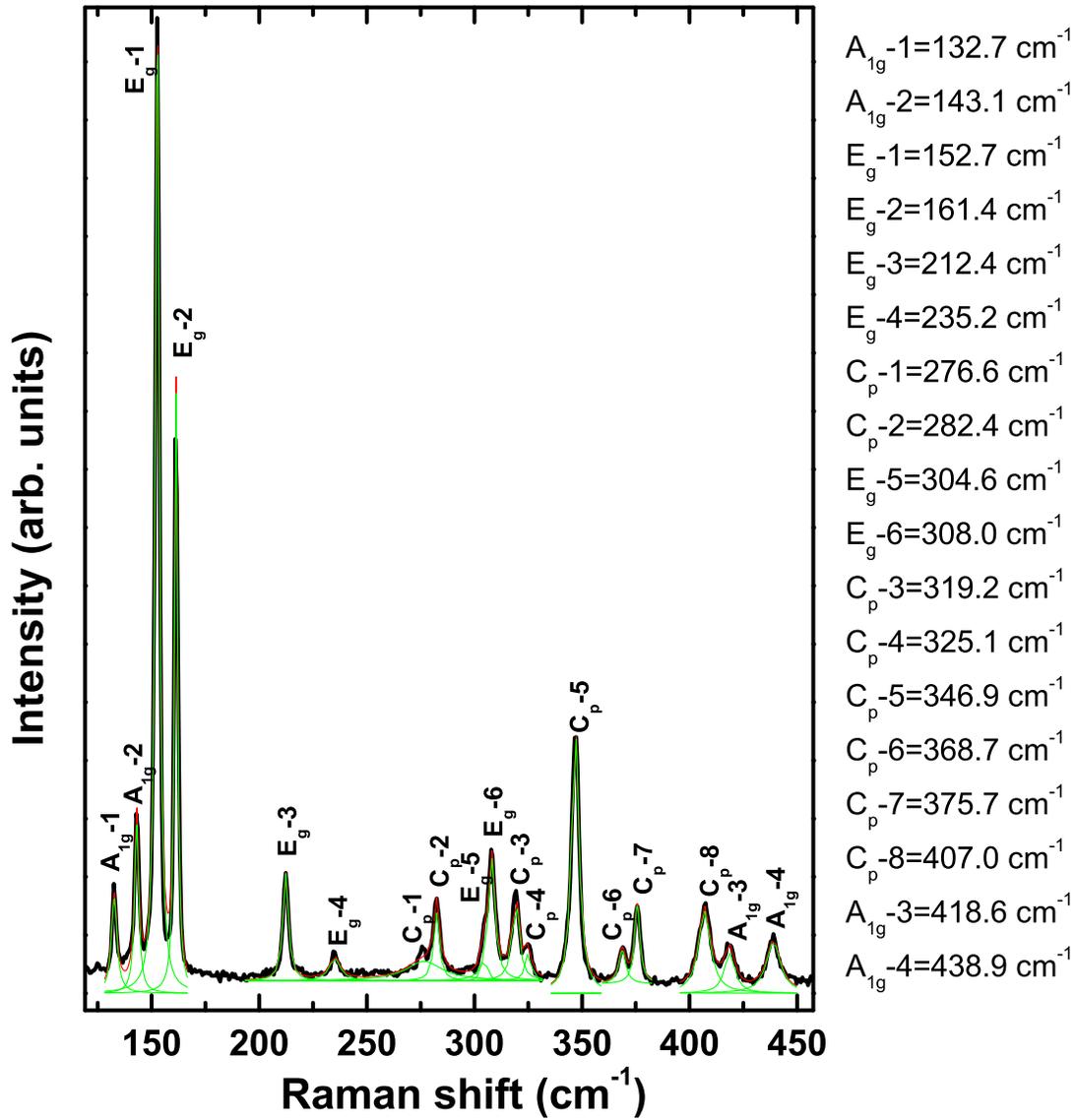}
	\caption{ Raman spectrum of distorted 1T $ReS_2$ placed on the lower diamond anvil. All the mode values are indicated in the vertical list.}
\end{figure}
\newpage

\begin{figure}[h!]
	\centering
	\includegraphics[width=1.0\textwidth]{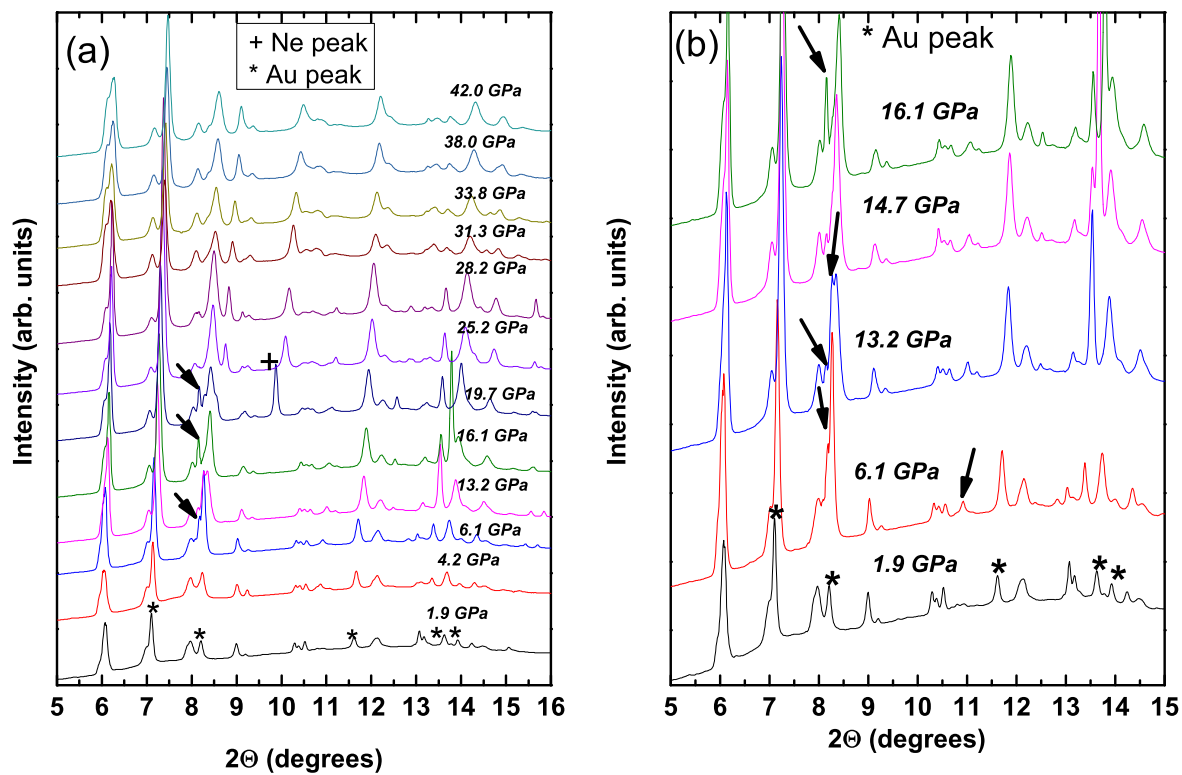}
	\caption{ (a) The evolution of the X-ray diffraction patterns at selected pressure points. (b) X-ray diffraction patters at selected pressure points in the range 1.9-16.1 GPa to show new peaks clearly.}
\end{figure}
\newpage
\begin{figure}[h!]
	\centering
	\includegraphics[width=0.9\textwidth]{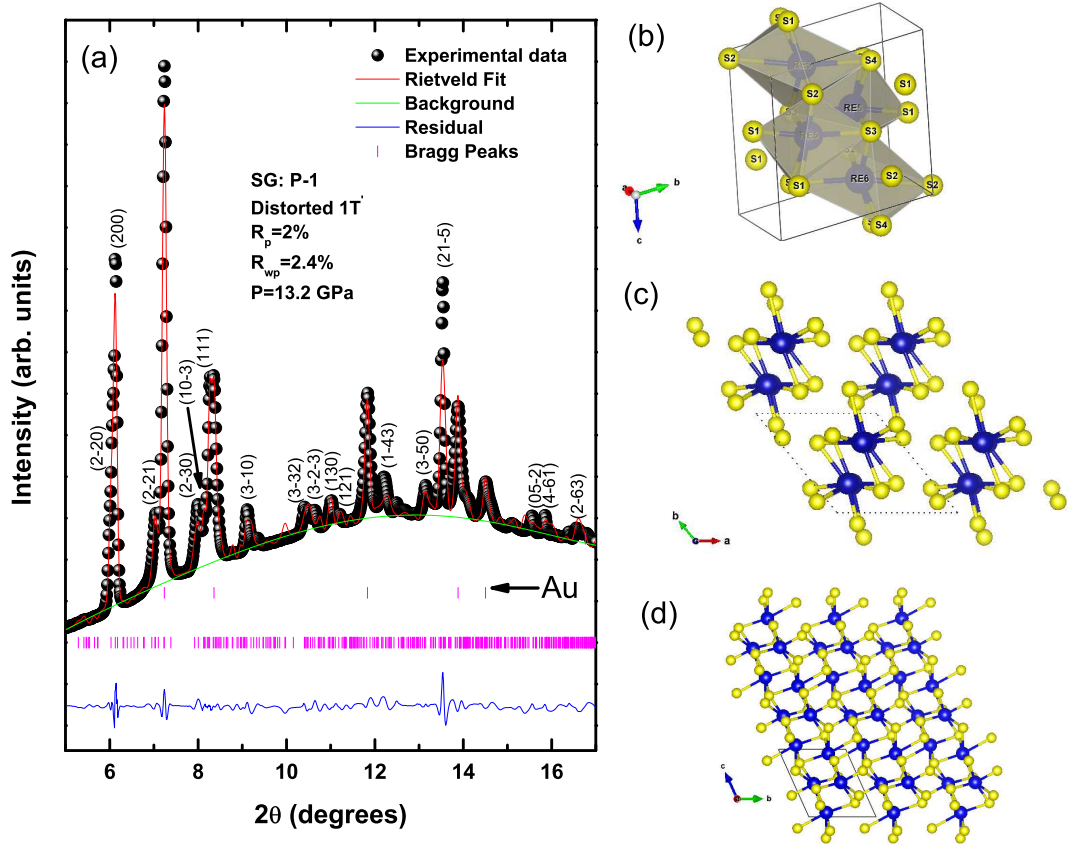}
	\caption{ (a) Rietveld fitting of the X-ray diffraction pattern of distorted 1T$^{\prime}$ $ReS_2$ sample at 13.2 GPa. (b) The view of the octahedra in a unit cell of distorted 1T$^{\prime}$ $ReS_2$. (c) The cross sectional view of layered 1T$^{\prime}$ $ReS_2$. (d) Top view of the mono-layered distorted 1T$^{\prime}$ $ReS_2$.}
\end{figure}

\begin{figure}[h!]
	\centering
	\includegraphics[width=0.9\textwidth]{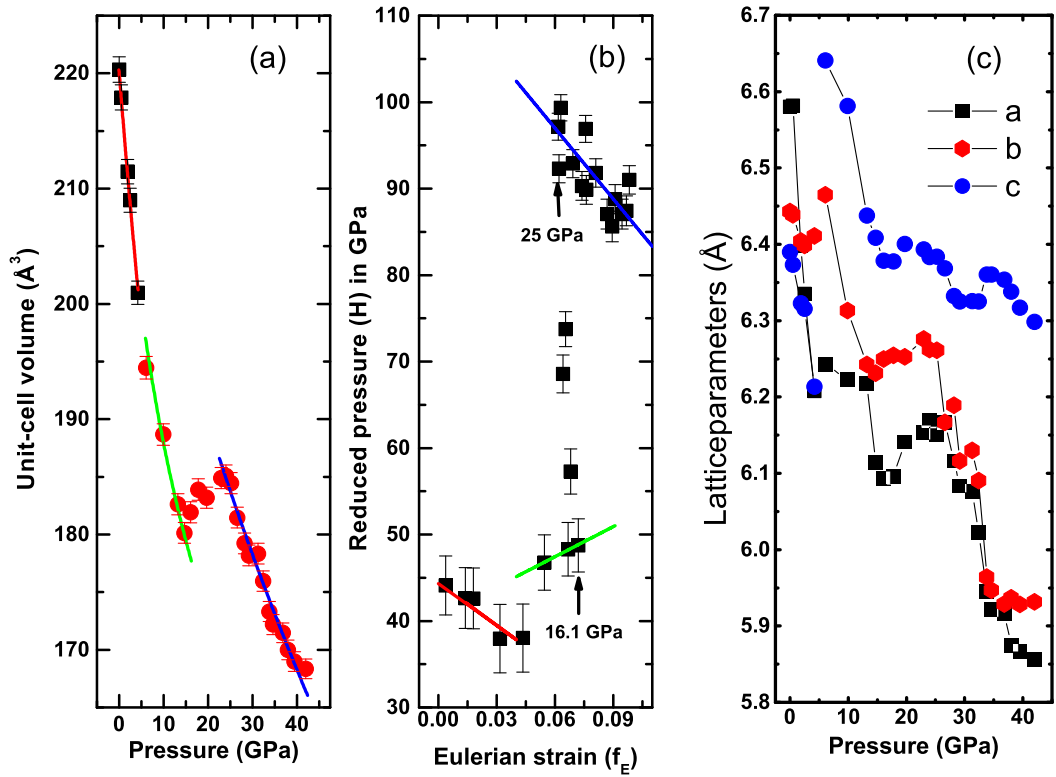}
	\caption{ (a) Evolution of volume as a function of pressure with EOS fitting (solid lines). Black filled squares represent volume of distorted 1T-phase, and red filled circles represent volume of distorted 1T$^{\prime}$-phase of $ReS_2$. (b) Eulerian strain ($f_E$) versus reduced pressure ($H$) in the entire pressure range of our study. (c) Evolution of lattice parameters with pressure in both phases of $ReS_2$.}
\end{figure}

\newpage
\begin{figure}[h!]
	\centering
	\includegraphics[width=1.0\textwidth]{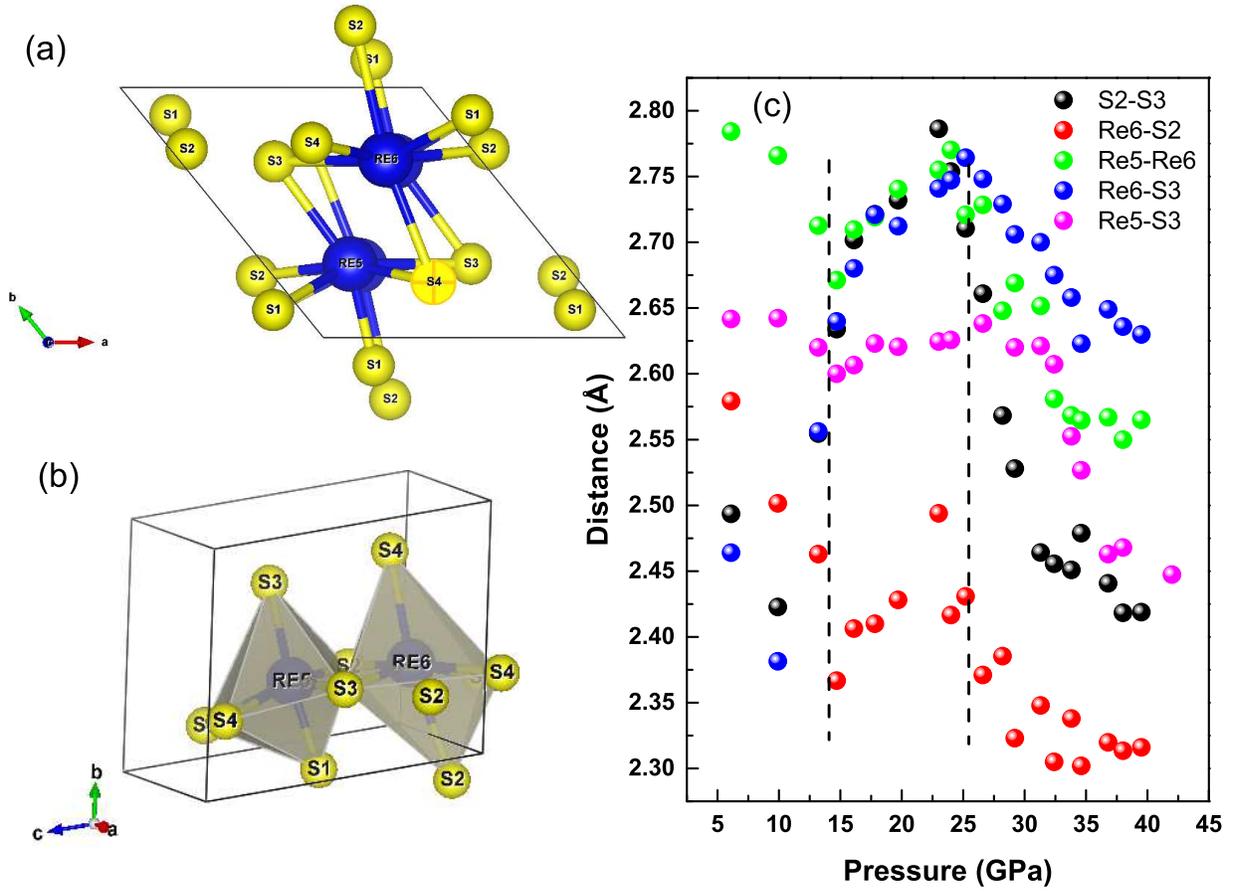}
	\caption{ (a) Vesta plot of the unit cell of distorted 1T$^{\prime}$-phase in $ab$-plane. (b) Two types of octahedra sharing S3 atom. (c) Evolution of atom distances as a function of pressure.}
\end{figure}

\newpage
\begin{figure}[h!]
	\centering
	\includegraphics[width=1.0\textwidth]{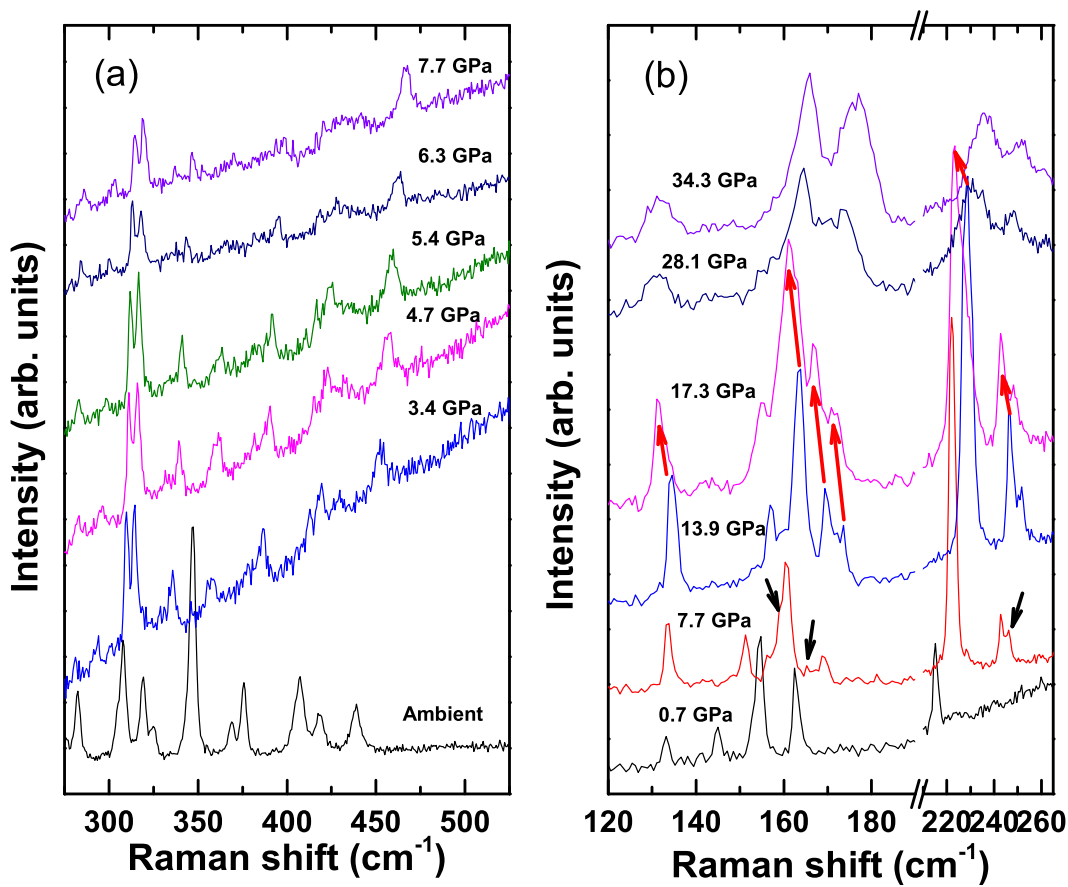}
	\caption{ (a) Evolution of Raman spectrum in the low pressure range 0-7.7 GPa from 250 to 550 $cm^{-1}$ at selected pressure points. (b) Evolution of Raman spectrum at the pressure range 0.7-34.3 GPa from 120 to 265 $cm^{-1}$ at selected pressure points. Black arrows represent new modes, while red arrows show the mode softening.}
\end{figure}

\newpage
\begin{figure}[h!]
	\centering
	\includegraphics[width=1.1\textwidth]{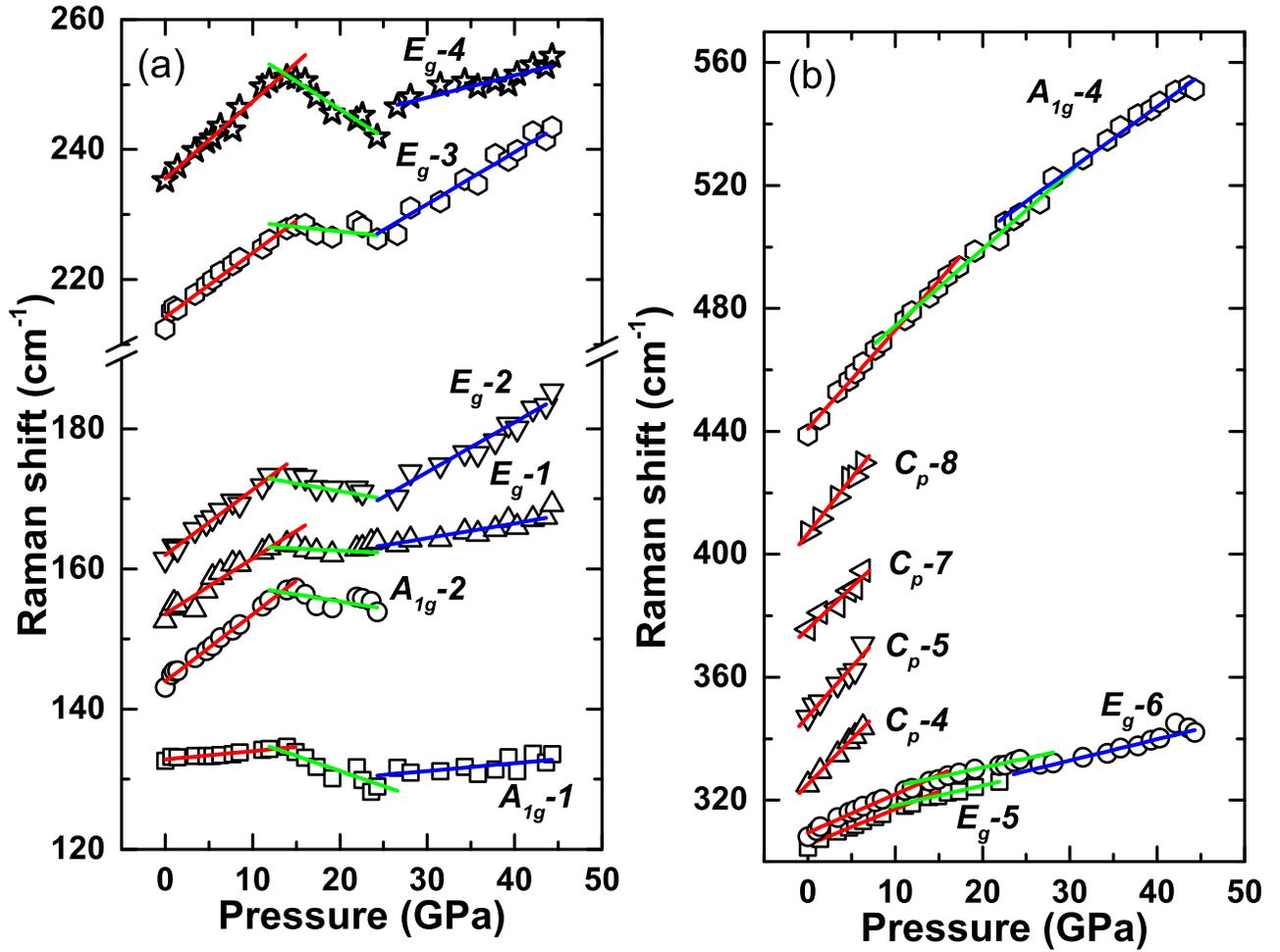}
	\caption{ (a) Pressure evolution of Raman modes corresponding to the rhenium atom vibrations.  (b) Raman modes corresponding to the sulfur atom vibrations with pressure. Red, green and blue lines represent linear fitting to the data in the pressure rang 0-14, 15-25, and 25-45 GPa, respectively.}
\end{figure}

\begin{figure}[h!]
	\centering
	\includegraphics[width=0.8\textwidth]{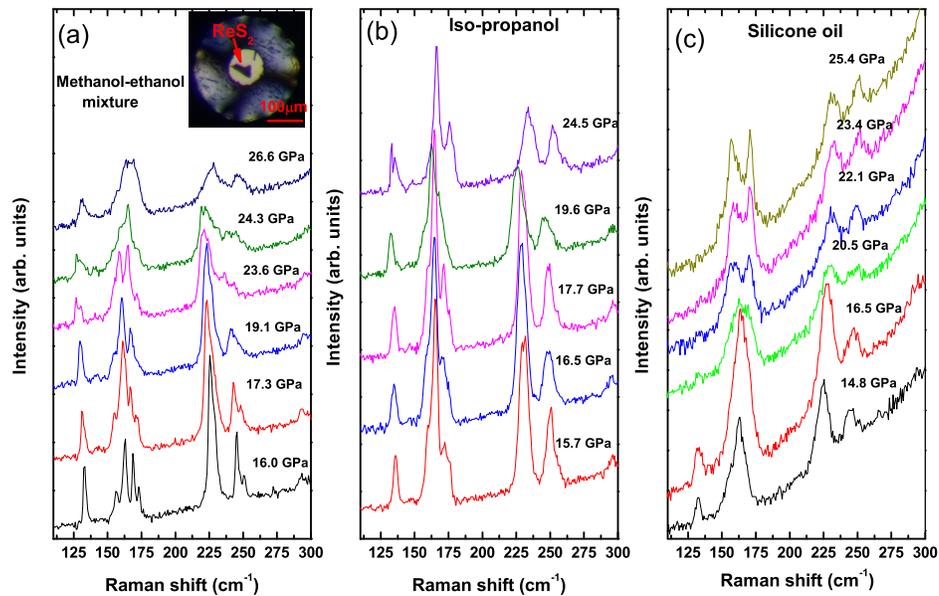}
	\caption{ A few Raman spectra with respect to pressure from 14.8-26.6 GPa are compared: (a) using ethanol-methanol mixture, (b) iso-propanol, (c) silicone oil. The inset of (a) depicts loaded DAC with $ReS_2$ at 17 GPa.}
\end{figure}

\end{document}